\documentclass[12pt]{article}

\usepackage{graphics,amsmath,amsthm,makeidx}
\usepackage{amssymb}  
\usepackage{amsfonts}
\usepackage{graphicx}
\usepackage{float}

\usepackage{times} 

\title{\Large \bf
Discrete-Time Models Resulting From\\
Dynamic Continuous-Time Perturbations In\\
Phase-Amplitude Modulation-Demodulation Schemes*
}

\author{Omer Tanovic$^{1}$, Alexandre Megretski$^{1}$, Yan Li$^{2}$, 
Vladimir M. Stojanovic$^{3}$, and
Mitra Osqui$^{4}$
\thanks{*This work was supported by 
DARPA Award No. W911NF-10-1-0088.}
\thanks{$^{1}$Omer Tanovic and Alexandre Megretski are with the Laboratory for Information and Decision Systems, Department of Electrical Engineering and Computer Science, Massachusetts Institute of Technology, Cambridge, MA 02139, USA
        {\tt\small \{otanovic,ameg\}@mit.edu}}%
\thanks{$^{2}$Yan Li was with the Laboratory for Information and Decision Systems, Department of Electrical Engineering and Computer Science, Massachusetts Institute of Technology. Currently she is with
NanoSemi Inc., Waltham, MA 02451, USA
        {\tt\small yan.li@nanosemitech.com}}%
\thanks{$^{3}$Vladimir M. Stojanovic is with the Department of Electrical Engineering and Computer Sciences, University of California Berkeley,
        Berkeley, CA 94720, USA
        {\tt\small vlada@berkeley.edu}}%
\thanks{$^{4}$Mitra Osqui was with the Laboratory for Information and Decision Systems, Department of Electrical Engineering and Computer Science, Massachusetts Institute of Technology. Currently she is a Research Scientist at Lyric Labs | Analog Devices, Cambridge, MA 02142, USA
        {\tt\small mitra.osqui@analog.com}}%
}
\begin{document}

\newcommand{\MRK}[1]{{\bf[#1]}}
\newcommand{\rf}[1]{(\ref{#1})}

\maketitle
\thispagestyle{empty}
\pagestyle{empty}

\begin{abstract}

We consider discrete-time (DT) systems S in which a
DT input is first tranformed to a continuous-time (CT) format by phase-amplitude
modulation, then modified by a non-linear CT dynamical transformation
F, and
finally converted back to DT output using an ideal de-modulation scheme.
Assuming that F belongs to a special class of  CT
Volterra series models with fixed
degree and memory depth, we provide a 
complete characterization of S as a series connection of
a DT Volterra series model of fixed degree and memory depth, and an LTI system with 
special properties. The result suggests a new, non-obvious, analytically motivated
 structure of digital compensation
of analog nonlinear distortions (for example, those caused by power amplifiers) in 
digital communication systems. 
Results from a MATLAB simulation are used to demonstrate
effectiveness of the new compensation scheme, as compared to
the standard Volterra series approach. 

\end{abstract}
\vskip5mm

\section{Notation and Terminology}
\begin{tabular}{ll}
$j$ & a fixed square root of $-1$\\
$\mathbb R$ & real numbers\\
$\mathbb Z$ & integers\\
$\mathbb N$ & positive integers\\
$[a:b]$ & all integers from $a$ to $b$\\
$\mathcal L$ & bounded square integrable functions $\mathbb R\to\mathbb R$\\
$\ell(X)$ & square summable functions $\mathbb Z\to X\subset\mathbb C^n$
\end{tabular}
\vskip3mm
\noindent{\sl CT signals} 
are elements of $\mathcal L$, {\sl DT signals} are elements of $\ell(X)$
for some $X\subset\mathbb C^n$. For $w\in\ell(X)$, $w[n]$ denotes
the value of $w$ at $n\in\mathbb Z$. 
In contrast, $x(t)$ refers to the value of $x\in\mathcal L$ at $t\in\mathbb R$.
{\sl Systems} are viewed as functions $\mathcal L\to\mathcal L$,
$\mathcal L\to\ell(X)$, $\ell(X)\to\mathcal L$, or $\ell(X)\to\ell(Y)$. $\mathbf Gf$
denotes the response 
of system $\mathbf G$ to signal $f$ (even when $\mathbf G$ is not linear), and the {\sl series
composition} $\mathbf K=\mathbf{QG}$ of systems
 $\mathbf Q$ and $\mathbf G$ is the system mapping $f$ to
$\mathbf{Q}(\mathbf{G}f)$.

\section{Introduction and Motivation}
Digital compensation offers an attractive approach to designing electronic 
devices
with superior characteristics \cite{Ken2000,Cripps2002,VuoRah2003}.
In this paper, a digital compensator is viewed as a system 
$\mathbf C:~\ell(\mathbb R)\to\ell(\mathbb R)$. More specifically,
a {\sl pre-}compensator $\mathbf C:~\ell(\mathbb R)\to\ell(\mathbb R)$
designed for a device modeled by a system 
$\mathbf P:~\ell(\mathbb R)\to\mathcal L$ (or
$\mathbf P:~\ell(\mathbb R)\to\ell(\mathbb R)$) aims to
make the composition $\mathbf{PC}$, as shown on the block diagram below,
\vskip3mm
\setlength{\unitlength}{0.15cm}
\begin{center}\begin{picture}(64,8)(14,50)
\put(26,50){\framebox(12,8){$ \mathbf C$}}
\put(52,50){\framebox(12,8){$ \mathbf P$}}
\put(14,54){\vector(1,0){12}}
\put(38,54){\vector(1,0){14}}
\put(64,54){\vector(1,0){12}}
\put(14,56){$u$}
\put(44,56){$w$}
\put(76,56){$v$}
\end{picture}\end{center}
\noindent conform to a set of desired specifications.
(In the simplest scenario, the objective is to make $\mathbf{PC}$
as close to the identity map as possible,
in order to cancel the distortions introduced by 
$\mathbf P$.)

A common element in digital compensator design algorithms is
selection of {\sl compensator structure}, which usually means
specifying a finite sequence $\tilde{\mathbf C}=(\mathbf C_1,\dots\mathbf C_N)$
of systems $\mathbf C_i:~\ell(\mathbb R)\to\ell(\mathbb R)$, and restricting the
actual compensator $\mathbf C$ to have the form
\[ \mathbf C=\sum_{i=1}^N a_i\mathbf C_i,\qquad a_i\in\mathbb R,\]
i.e., to be a linear combination of the elements of $\tilde{\mathbf C}$.
Once the {\sl basis} sequence $\tilde{\mathbf C}$ is fixed, the design usually reduces to
a straightforward {\sl least squares optimization} of the coefficients $a_i\in\mathbb R$.

A popular choice is for the systems $\mathbf C_k$ to be some {\sl Volterra monomials},
i.e. to map their input $u=u[n]$ to the outputs $w_k=w_k[n]$ according to the polynomial 
formulae
\[  w_k[n] = \prod_{i=1}^{i=d_k}u[n-n_{k,i}]\]
(where the integers $d_k$, $n_{k,i}$ will be referred to,
respectively, as the {\sl degrees} and {\sl delays}),
which makes every linear combination $\mathbf C$ of $\mathbf C_i$ a
{\sl DT Volterra series}  \cite{Schetz2006}, i.e., a  DT system
mapping signal inputs $u\in\ell(\mathbb R)$ to outputs $w\in\ell(\mathbb R)$ 
according to the polynomial expression
\[ w[n]=\sum_{k=1}^{N}a_k\prod_{i=1}^{d_k}u[n-n_{k,i}].\]

Selecting a proper {\sl compensator structure} is a major challenge
in compensator design: a basis
 which is too simple will not be capable of cancelling the distortions
well, while a form that is too complex will consume excessive power
and space. Having an insight into the compensator
basis selection can be very valuable. For  an example (cooked up outrageously
to make the point), consider the case when the ideal 
compensator $\mathbf C:~u\mapsto w$ is
given by
\[  w[n] = \rho u[n]+\delta\left(\sum_{j=-50}^{50}u[n-j]\right)^5\]
for some (unknown) coefficients $\rho$ and $\delta$.
One can treat $\mathbf C$ as a generic Volterra series expansion
with fifth order monomials with delays between $-50$ and $50$, and the first order monomial
with delay 0, which leads to a basis sequence $\tilde{\mathbf C}$ with
$1+{105\choose 5}=96560647$ elements (and the same number of multiplications involved
in implementing the compensator). Alternatively, one may realize that
the two-element structure $\tilde{\mathbf C}=\{\mathbf C_1,\mathbf C_2\}$, with
$w_i=\mathbf C_i u$ defined by
\[  w_1[n]=u[n],\qquad w_2[n]=\left(\sum_{j=-50}^{50}u[n-j]\right)^5\]
is good enough.
 
In this paper we establish that a certain special 
structure is good enough to compensate for
imperfect modulation. We consider systems 
represented by the block diagram

\setlength{\unitlength}{0.12cm}
\begin{center}\begin{picture}(84,15)(8,64)
\put(24,64){\framebox(16,12){$ \mathbf M$}}
\put(58,64){\framebox(16,12){$ \mathbf F$}}
\put(8,70){\vector(1,0){16}}
\put(40,70){\vector(1,0){18}}
\put(74,70){\vector(1,0){18}}
\put(8,72){$u[n]$}
\put(48,72){$x(t)$}
\put(92,72){$y(t)$}
\end{picture}\end{center}
where $\mathbf M:~\ell(\mathbb C)\to\mathcal L$ is the 
{\sl ideal} modulator with fixed sampling interval length $T>0$
and modulation-to-sampling frequency ratio $M\in\mathbb N$, 
converting complex DT signals $w\in\ell(\mathbb C)$ 
 to CT signals $x\in\mathcal L$ according to
\begin{equation}\label{eq:mod}
  x(t)=\sum_{n\in\mathbb Z}~
\frac{1}{T}~p\left(\frac{t}T-n\right)\text{Re}\left\{\exp\left(j\frac{2\pi M}{T}t\right)u[n]\right\},
\end{equation}
with
\[  p(t)=\begin{cases}1,& t\in[0,1),\\ 0,& t\not\in[0,1),\end{cases}\]
and $\mathbf F:~\mathcal L\to\mathcal L$ 
is a CT dynamical system used to represent linear and 
nonlinear distortion in the modulator and power amplifier circuits. In particular,
we are interested in the case where the relation between $x(\cdot)$ and $y(\cdot)$
is described by the {\sl CT Volterra series model}
\begin{equation}\label{eq:ctvs}
  y(t)=b_0+\sum_{k=1}^{N_b}b_k\prod_{i=1}^{\beta_k}x(t-t_{k,i}),
\end{equation}
where $N_b\in\mathbb N$, $b_k\in\mathbb R$, $\beta_k\in\mathbb N$, $t_{k,i}\ge0$
are parameters. (In a similar fashion,  it is possible to consider input-output relations
in which the finite sum in (\ref{eq:ctvs}) is replaced by an integral, or an infinite sum).
One expects that the memory of $F$ is not long, compared to $T$,
i.e., that $\max t_{k,i}/T$ is not much larger than 1.

As a rule, the spectrum of the DT input $u\in\ell(\mathbb C)$ of the modulator is carefully 
shaped at a pre-processing stage
to guarantee desired characteristics of the modulated signal $x=\mathbf Mu$. However,
when the distortion $\mathbf F$ is not linear, the spectrum of the
$y=\mathbf Fx$ could be damaged substantially, 
leading to violations of  EVM and spectral mask
specifications \cite{And_et_all2008}.

Consider the possibility of repairing the spectrum of $y$ by pre-distorting the digital input
$u\in\ell(\mathbb C)$ by a compensator 
$\mathbf C:~\ell(\mathbb C)\to\ell(\mathbb C)$, 
as shown
on the block diagram below:
\setlength{\unitlength}{0.16cm}
\begin{center}\begin{picture}(72,10)(4,60)
\put(14,60){\framebox(12,8){$ \mathbf C$}}
\put(34,60){\framebox(12,8){$ \mathbf M$}}
\put(54,60){\framebox(12,8){$\mathbf F$}}
\put(4,64){\vector(1,0){10}}
\put(26,64){\vector(1,0){8}}
\put(46,64){\vector(1,0){8}}
\put(66,64){\vector(1,0){10}}
\put(48,66){$x(t)$}
\put(72,66){$y(t)$}
\put(4,66){$u[n]$}
\put(28,66){$w[n]$}
\end{picture}\end{center}
The desired effect of inserting $\mathbf C$ 
is cancellation of the distortion caused by 
$\mathbf F:~\mathcal L\to\mathcal L$.
Naturally, since $\mathbf C$ acts in the baseband (i.e., in discrete time), 
there is no chance that
$\mathbf C$ will achieve a complete correction, i.e., that the series composition 
$\mathbf F\mathbf M\mathbf C$ of $\mathbf F$, 
$\mathbf M$, and $\mathbf C$ will
be identical to $\mathbf M$. 
However, in principle, it is sometimes possible to make the frequency
contents of $\mathbf Mu$ and $\mathbf F\mathbf M\mathbf Cu$ to be identical 
within the CT frequency band $(f_c-f_N,f_c+f_N)$ Hz, where
$f_c=M/T$ is the carrier frequency (Hz), and
$f_N=0.5/T$ is the Nyquist frequency (Hz) for the sampling rate used
\cite{Tsimbinos1998}.
To this end, let $\mathbf H:~\mathcal L\to\mathcal L$ 
denote the ideal band-pass filter with frequency response
\[  H(f)=\begin{cases}1,& |f|\in(f_c-f_N,f_c+f_N),\\ 0,& |f|\not\in(f_c-f_N,f_c+f_N).
\end{cases}\]
Let $\mathbf D:~\mathcal L\to\ell(\mathbb C)$ 
be the ideal de-modulator relying on the band selected by 
$\mathbf H$, i.e. the linear system for which the
series composition $\mathbf D\mathbf H\mathbf M$ is the identity
function. Let 
$\mathbf S=\mathbf D\mathbf H\mathbf F\mathbf M$ be the series composition of
$\mathbf D$, $\mathbf H$, $\mathbf F$, and $\mathbf M$, i.e. the 
DT system
with input $w=w[n]$ and output $v=v[n]$ shown on the block diagram below:
\begin{figure}[H]
\setlength{\unitlength}{0.16cm}
\begin{center}\begin{picture}(96,11)(2,58)
\put(12,58){\framebox(12,8){$\mathbf M$}}
\put(34,58){\framebox(12,8){$\mathbf  F$}}
\put(56,58){\framebox(12,8){$\mathbf  H$}}
\put(78,58){\framebox(10,8){$\mathbf  D$}}
\put(2,62){\vector(1,0){10}}
\put(24,62){\vector(1,0){10}}
\put(46,62){\vector(1,0){10}}
\put(68,62){\vector(1,0){10}}
\put(88,62){\vector(1,0){10}}
\put(2,64){$w[n]$}
\put(25,64){$x(t)$}
\put(47,64){$y(t)$}
\put(69,64){$q(t)$}
\put(94,64){$v[n]$}
\end{picture}\end{center}
\caption{}
\label{fig:whole_chain}
\end{figure}
By construction, the ideal compensator $\mathbf C$ should be the inverse
$\mathbf C=\mathbf S^{-1}$ of $\mathbf S$, as long as the inverse does exist.

A key question answered in this paper is "what to expect from system $\mathbf S$?"
If one assumes that the continuous-time distortion subsystem $\mathbf F$ is
simple enough, what does this say about $\mathbf S$?

This paper provides an explicit expression for $\mathbf S$ in the case when
$\mathbf F$ is given in the CT Volterra series form \rf{eq:ctvs} with
{\sl degree} $d=\max\beta_k$ and {\sl depth} $t_{max}=\max t_{k,i}$. 
The result reveals that, even though $\mathbf S$ tends to have infinitely long memory
(due to the ideal band-pass filter $\mathbf H$ being involved in the construction of
$\mathbf S$), it can be represented as a series composition 
$\mathbf S=\mathbf L\mathbf V$, where  
$\mathbf V:~\ell(\mathbb C)\to\ell(\mathbb R^N)$ maps scalar
complex input $w\in\ell(\mathbb C)$ to real vector output 
$g\in\ell(\mathbb R^N)$ in such a way that
the $k$-th scalar component $g_k[n]$ of $g[n]\in\mathbb R^N$
is given by
\[  g_k[n]=\prod_{i=0}^{m}(\text{Re}~w[n-i])^{\alpha_i}~
\prod_{i=0}^{m}(\text{Im}~w[n-i])^{\beta_i},\]
\[\alpha_i,\beta_i\in\mathbb Z_+,\qquad
 \sum_{i=0}^m\alpha_i+\sum_{i=0}^m\beta_i\le d,\]
$m$ is the minimal integer not smaller than  $t_{max}/T$,
and  $\mathbf L:~\ell(\mathbb R^N)\to\ell(\mathbb C)$
is an LTI system.  
\begin{figure}[H]
\setlength{\unitlength}{0.115cm}
\begin{center}\begin{picture}(84,15)(8,64)
\put(24,64){\framebox(16,12){$ \mathbf V$}}
\put(58,64){\framebox(16,12){$ \mathbf L$}}
\put(8,70){\vector(1,0){16}}
\put(40,70){\vector(1,0){18}}
\put(74,70){\vector(1,0){18}}
\put(8,72){$w[n]$}
\put(48,72){$g[n]$}
\put(92,72){$v[n]$}
\end{picture}\end{center}
\caption{Block diagram of the structure of {\bf S}}
\label{fig:compensator}
\end{figure}
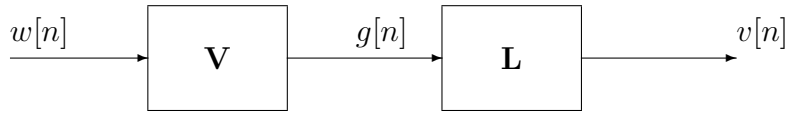
Moreover, $\mathbf L$ can be shown to have a good approximation of the form
$\mathbf L\approx X\mathbf L_0$, where $X$ is a static gain matrix, and $\mathbf L_0$
is an LTI model which does not depend on $b_k$ and $t_{k,i}$.
In other words, $\mathbf S$ can be well approximated by combining a Volterra series model
with a short memory, and a {\sl fixed} (long memory) LTI, as long as
the memory depth $t_{max}$ of $\mathbf F$ is short, relative to the sampling time $T$.

In most applications, with an appropriate scaling and time delay, the system $\mathbf S$
to be inverted can be viewed as a small perturbation of identity, i.e. 
 $\mathbf S=\mathbf I+\mathbf\Delta$. When $\mathbf\Delta$ is  "small" in an appropriate
sense (e.g., has small incremental L2 gain $\|\mathbf\Delta\|\ll1$), the inverse of $\mathbf S$
can be well approximated by $\mathbf S^{-1}\approx\mathbf I-\mathbf\Delta=2\mathbf I-\mathbf S$.
Hence the result of this paper suggests a specific structure of the 
compensator (pre-distorter) $\mathbf C\approx\mathbf I-\mathbf\Delta=2\mathbf I-\mathbf S$. 
In other words, a plain Volterra monomials structure is, in general,  not good enough for $\mathbf C$, 
as it lacks the capacity to implement the long-memory
LTI post-filter $\mathbf L$. Instead, $\mathbf C$ should be sought
in the form $\mathbf C=\mathbf I-\mathbf{L_0}X\mathbf{V}$, where $\mathbf V$ is the system
generating all Volterra series monomials of a limited depth and limited degree,
$L_0$ is a {\sl fixed} LTI system with a very long time constant, and $X$ is a matrix of coefficients
to be optimized to fit the data available.

\section{Main Result}

Given a sequence $\tau=(\tau_1,\dots,\tau_d)$ of $d$ non-negative real numbers
$\tau_i$ let $\mathbf F_{\tau}:~\mathcal{L}\to \mathcal{L}$ be the CT system 
mapping inputs $x\in\mathcal L$ to the outputs $y\in\mathcal L$ defined by
\[y(t)=x(t-\tau_1)x(t-\tau_2)\dots x(t-\tau_d).\] 
Given d-tuple $m=(m_1,\dots,m_d)\in [1:4]^d$ 
and $k\in[1:4]$ let $S_k(m)=\{i\in [1:d]:~m_i=k\}$, and define
\[ N_1(m)=|S_1(m)\cup S_2(m)|,~ N_2(m)=|S_3(m)\cup S_4(m)|.\]

Let $(\cdot ,\cdot):\mathbb{R}^d\times \mathbb{R}^d \to \mathbb{R}$ be a map 
defined by $(x,y)=\sum_{i=1}^d x_i\cdot y_i$ (i.e. the standard scalar product in $\mathbb{R}^d$), 
and let maps $\bar{\sigma},\sigma:\mathbb{R}^d\to \mathbb{R}$ be defined by 
$\bar{\sigma}(x)=\sum_{i=1}^d x_i$ and $\sigma(x)=\bar{\sigma}(x)-1$. Also for 
a given $m\in [1:4]^d$, we define map $\pi_m:\mathbb{R}^d\to \mathbb{R}$ by 
\[\pi_m(x)=\prod_{i\in S_3(m)\cup S_4(m)}x_i,\] 
and projection operators $\mathcal{P}_{m}^{i}:\mathbb{R}^d\to \mathbb{R}^{N_i(m)}, i=1,2$ by  \[\mathcal{P}_{m}^{i}x=\begin{bmatrix} x_{n_1}&\dots&x_{n_{N_i(m)}}\end{bmatrix}^T,\] 
\[\{n_1, \dots, n_{N_i(m)}\}=S_i(m)\cup S_{i+1}(m), n_1<\dots<n_{N_i(m)}.\]


\begin{figure}[H]
\footnotesize
\setlength{\unitlength}{0.16cm}
\begin{center}\begin{picture}(88,21)(0,37)
\put(0,52){\vector(1,0){8}}
\put(8,48){\framebox(8,8){$ ZOH$}}
\put(28,48){\framebox(8,8){$ Re\{\cdot\}$}}
\put(36,52){\vector(1,0){6}}
\put(42,48){\framebox(8,8){$ F_\tau(\cdot)$}}
\put(62,48){\framebox(8,8){$ LPF$}}
\put(70,52){\line(1,0){6}}
\put(76,52){\line(1,1){4}}
\put(80,52){\vector(1,0){8}}
\put(22,52){\circle{6}}
\put(56,52){\circle{6}}
\put(22,38){\vector(0,1){11}}
\put(56,38){\vector(0,1){11}}
\put(24,38){$e^{j\omega_ct}$}
\put(58,38){$e^{-j\omega_ct}$}
\put(16,52){\vector(1,0){3}}
\put(25,52){\vector(1,0){3}}
\put(50,52){\vector(1,0){3}}
\put(59,52){\vector(1,0){3}}
\put(0,54){$w[n]$}
\put(86,54){$v[n]$}
\put(20,50){\line(1,1){4}}
\put(20,54){\line(1,-1){4}}
\put(54,50){\line(1,1){4}}
\put(54,54){\line(1,-1){4}}
\put(12,44){$T$}
\put(78,52){\vector(0,-1){2}}
\qbezier(76,58)(78,58)(78,52)
\put(78,44){$T$}
\put(52,56){$y(t)$}
\end{picture}\end{center}
\caption{\ \label{fig:Figure11}}
\end{figure}

Given a vector $\tau\in \mathbb{R}_+^d$ let $k$ be a vector in $(\mathbb{N}\cup \{0\})^d$, 
such that $\tau=kT+\tau'$, with $\tau'\in [0,T)^d$. It is obvious that 
for a given $\tau$ vector $k$ is uniquely defined.

Given a positive real number $T$, let us denote by $p_{ZOH}(t)$, 
impulse response of the zero-order hold (ZOH) system. 
We have $p_{ZOH}(t)= \frac{1}{T}(u(t)-u(t-T))$, where $u(t)$ is 
the Heaviside step function. Moreover for a given $m\in[1:4]^d$ 
and $\tau'\in [0,T)^d$,we define 
\[\tau_{min}=\begin{cases}\min_{i\in S_2(m)\cup S_4(m)} \tau'_i,&
|S_2(m)\cup S_4(m)|>0\\0,&o/w \end{cases},\] 
and 
\[\tau_{max}=\begin{cases} \max_{i\in S_1(m)\cup S_3(m)}\tau'_i,&
|S_1(m)\cup S_3(m)|>0\\T,&o/w \end{cases}.\]
Now let $p_{m,\tau}:\mathbb{R}\to \mathbb{R}$ be the continuous time signal
 defined by 
\[p_{m,\tau}(t)=\begin{cases}\frac{1}{T}(u(t-\tau_{min})-u(t-\tau_{max})),& 
\tau_{min}<\tau_{max}\\0,&o/w \end{cases}.\] 
We denote its Fourier transform by $P_{m,\tau}(j\omega)$.

From (2) we can see that general CT Volterra model is a 
linear combination of subsistems of form $F_{\tau}$, so in order 
to find system decomposition $\mathbf {S=LV}$  it is clearly sufficient 
to find what happens with one particular element $F_{\tau}$, i.e. to 
find map $\mathbf{DHF_{\tau}M}$. The following theorem gives 
answer to that question.
\vskip2mm
\textbf{Theorem 2.1.} 
A DT system $\mathbf{DHF_{\tau}M}:\ell(\mathbb{C})\to \ell(\mathbb{C})$, 
mapping $w[n]=i[n]+j\cdot q[n]$ to $v[n]$, is given by
\[v[n]=\sum_{m\in \{1,2,3,4\}^d}f_{m,k}[n]*h_{m,\tau}[n],\]
where 
\[f_{m,k}[n]=\prod_{i\in S_1(m)}i[n-k_i-1] \cdot \prod_{i\in S_2(m)} i[n-k_i]
\cdot \prod_{i\in S_3(m)} q[n-k_i-1] \cdot \prod_{i\in S_4(m)}q[n-k_i],\]
and Fourier transform of a unit sample response $h_{m,\tau}[n]$ is 
given by
\[H_{m,\tau}(e^{j\Omega})=\frac{ (-j)^{N_2(m)}}{2^d} \sum_{r\in\{-1,1\}^d}(-1)^{\pi_m(r)}
P_{m,\tau}\left(j\frac{\Omega}{T}-j\omega_c \sigma(r)\right)  e^{-j\omega_c(r,\tau)}.\]
\vskip3mm
\begin{proof}
We first state and prove the following Lemma, 
which is very similar to Theorem 2.1 but considers 
somewhat simpler case when  $\tau\in [0,T)^d$, i.e. $k=\mathbf{0}$. 
The proof of Theorem 2.1 then immediately follows from this Lemma. 
\vskip3mm

\textbf{Lemma 2.2.} 
Suppose that $\tau\in [0,T)^d$. A DT system 
$\mathbf{DHF_{\tau}M}:\ell(\mathbb{C})\to \ell(\mathbb{C})$, 
mapping $w[n]=i[n]+j\cdot q[n]$ to $v[n]$, is given by
\[v[n]=\sum_{m\in \{1,2,3,4\}^d}f_{m}[n]*h_{m,\tau}[n],\]
where 
\[f_m[n]=i[n-1]^{|S_1(m)|} \cdot i[n]^{|S_2(m)|} \cdot 
q[n-1]^{|S_3(m)|} \cdot q[n]^{|S_4(m)|},\]
and Fourier transform of a unit sample response $h_{m,\tau}[n]$ is 
given by
\[ H_{m,\tau}(e^{j\Omega})=\frac{ (-j)^{N_2(m)}}{2^d} \sum_{r\in\{-1,1\}^d}(-1)^{\pi_m(r)}
P_{m,\tau}\left(j\frac{\Omega}{T}-j\omega_c \sigma(r)\right) e^{-j\omega_c(r,\tau)}.\]
\vskip3mm
\begin{proof}
Let us first analyze what happens in the case 
when $d=1$, i.e. system $\mathbf{F_{\tau}}$ is just a delay by $\tau\in[0,T)$. 
Output $y(t)$ of $F_{\tau}$ becomes 
\[ y(t)=i(t-\tau)\cos(t-\tau)-q(t-\tau)\sin(t-\tau). \]
We observe that $\mathbf{F_{\tau}}$ commutes with the modulation 
subsystem M, following an appropriate splitting of the ZOH 
impulse response, thus allowing us to move $\mathbf{F_{\tau}}$ out of the 
Mod/Demod part of the system. Now system $\mathbf{DHF_{\tau}M}$ is 
equivalent to the one shown in Fig.~\ref{fig:F_tauM}, where the impulse 
responses $p_1(t)$ and $p_2(t)$ are given by 
\[ p_1(t) =  \frac{1}{T}(u(t)-u(t-\tau)),\]
\[ p_2(t) =  \frac{1}{T}(u(t-\tau)-u(t-T)).\]

\begin{figure}[h]
\footnotesize
\setlength{\unitlength}{0.15cm}
\begin{center}\begin{picture}(101,72)(0,16)
\put(0,52){\vector(1,0){6}}
\put(0,54){$w[n]$}
\put(92,56){\line(-1,-1){4}}
\put(92,52){\vector(1,0){6}}
\put(90,52){\vector(0,-1){2}}
\qbezier(88,58)(90,58)(90,52)
\put(90,46){$T$}
\put(96,54){$v[n]$}
\put(78,48){\framebox(8,8){$ LPF$}}
\put(86,52){\line(1,0){2}}
\put(6,38){\framebox(3,28){$ $}}
\put(14,62){\line(-1,0){5}}
\put(14,62){\line(0,1){18}}
\put(14,80){\vector(1,0){4}}
\put(18,76){\framebox(6,8){$ z^{-1}$}}
\put(28,76){\framebox(8,8){$ p_1(t)$}}
\put(9,42){\vector(1,0){9}}
\put(14,42){\line(0,-1){18}}
\put(14,24){\vector(1,0){14}}
\put(28,20){\framebox(8,8){$ p_2(t)$}}
\put(18,38){\framebox(6,8){$ z^{-1}$}}
\put(24,80){\vector(1,0){4}}
\put(24,42){\vector(1,0){4}}
\put(28,38){\framebox(8,8){$ p_1(t)$}}
\put(14,62){\vector(1,0){14}}
\put(28,58){\framebox(8,8){$ p_2(t)$}}
\put(44,24){\circle{6}}
\put(36,24){\vector(1,0){5}}
\put(36,28){$e_4(t)$}
\put(44,42){\circle{6}}
\put(36,42){\vector(1,0){5}}
\put(44,62){\circle{6}}
\put(36,62){\vector(1,0){5}}
\put(44,80){\circle{6}}
\put(36,80){\vector(1,0){5}}
\put(36,46){$e_3(t)$}
\put(36,66){$e_2(t)$}
\put(36,84){$e_1(t)$}
\put(42,82){\line(1,-1){4}}
\put(42,78){\line(1,1){4}}
\put(42,60){\line(1,1){4}}
\put(42,64){\line(1,-1){4}}
\put(42,40){\line(1,1){4}}
\put(42,44){\line(1,-1){4}}
\put(42,26){\line(1,-1){4}}
\put(42,22){\line(1,1){4}}
\put(44,56){\vector(0,1){3}}
\put(44,74){\vector(0,1){3}}
\put(44,48){\vector(0,-1){3}}
\put(44,30){\vector(0,-1){3}}
\put(40,72){$cos(t-\tau)$}
\put(40,54){$cos(t-\tau)$}
\put(40,50){$sin(t-\tau)$}
\put(40,32){$sin(t-\tau)$}
\put(48,84){$f_1(t)$}
\put(48,66){$f_2(t)$}
\put(48,46){$f_3(t)$}
\put(48,28){$f_4(t)$}
\put(47,62){\line(1,0){7}}
\put(47,80){\line(1,0){7}}
\put(54,80){\vector(1,-4){6}}
\put(60,52){\circle{6}}
\put(47,42){\line(1,0){7}}
\put(54,42){\vector(1,2){4}}
\put(54,62){\vector(1,-2){4}}
\put(47,24){\line(1,0){7}}
\put(54,24){\vector(1,4){6}}
\put(4,16){\dashbox(60,72){$ $}}
\put(60,54){\line(0,-1){4}}
\put(58,52){\line(1,0){4}}
\put(72,52){\circle{6}}
\put(75,52){\vector(1,0){3}}
\put(63,52){\vector(1,0){6}}
\put(62,54){$y(t)$}
\put(70,50){\line(1,1){4}}
\put(70,54){\line(4,-5){4}}
\put(72,40){\vector(0,1){9}}
\put(70,36){$e^{-j\omega_ct}$}
\put(66,86){$\mathbf{F{\tau}M}$}
\put(44,30){\line(0,1){2}}
\put(44,48){\line(0,1){2}}
\put(9,66){$i[n]$}
\put(9,46){$q[n]$}
\end{picture}\end{center}
\caption{}
\label{fig:F_tauM}
\end{figure}

It is clear that $p_1(t)$ and $p_2(t)$ form the above mentioned 
splitting of the $p_{ZOH}(t)$, in the sense that the ZOH impulse 
response satisfies $p_{ZOH}(t)=p_1(t)+p_2(t)$. Thus subsystem 
$\mathbf{F_{\tau}M}$, mapping $w[n]$ to $y(t)$, can be represented 
as a parallel connection of four LTI systems whose inputs are current and 
previous values of in-phase and quadrature components of 
the input signal $w[n]$. Hence output $y(t)$ can be written as
\[y(t) = f_1(t)+f_2(t)+f_3(t)+f_4(t),\]
where definition of signals $f_i(t)$ is obvious from Fig.~\ref{fig:F_tauM}. 
\vskip3mm

\begin{figure}[h]
\footnotesize
\setlength{\unitlength}{0.12cm}
\begin{center}\begin{picture}(100,58)(0,34)
\put(0,86){\vector(1,0){16}}
\put(30,86){\line(1,0){10}}
\put(16,80){\framebox(14,12){$\mathbf{F{\tau_1}M}$}}
\put(16,62){\framebox(14,12){$\mathbf{F{\tau_2}M}$}}
\put(46,68){\circle{6}}
\put(40,40){\vector(1,4){6}}
\put(40,86){\vector(1,-3){5}}
\put(30,68){\vector(1,0){13}}
\put(30,40){\line(1,0){10}}
\put(16,34){\framebox(14,12){$\mathbf{F{\tau_d}M}$}}
\put(58,68){\circle{6}}
\put(61,68){\vector(1,0){5}}
\put(66,62){\framebox(14,12){$ LPF$}}
\put(84,68){\line(-1,0){4}}
\put(84,68){\line(1,1){6}}
\put(90,68){\vector(1,0){10}}
\put(98,70){$v[n]$}
\put(0,90){$w[n]$}
\put(44,66){\line(1,1){4}}
\put(44,70){\line(1,-1){4}}
\put(56,66){\line(1,1){4}}
\put(56,70){\line(1,-1){4}}
\put(58,56){\vector(0,1){9}}
\put(56,52){$e^{-j\omega_ct}$}
\put(49,68){\vector(1,0){6}}
\put(48,70){$y(t)$}
\put(87,68){\vector(0,-1){2}}
\qbezier(84,76)(87,76)(87,68)
\put(86,60){$T$}
\put(34,88){$f^1(t)$}
\put(34,72){$f^2(t)$}
\put(34,44){$f^d(t)$}
\put(10,86){\line(0,-1){46}}
\put(10,40){\vector(1,0){6}}
\put(10,68){\vector(1,0){6}}
\end{picture}\end{center}
\caption{}
\label{fig:F_tauMcompact}
\end{figure}

Now suppose that order $d$ of $\mathbf{F_{\tau}}$ is an arbitrary positive 
integer. From analysis in the case when $d=1$, it immediately 
follows that block structure shown in Fig.~\ref{fig:F_tauMcompact} is an equivalent 
representation of the system $\mathbf{DHF_{\tau}M}$. Hence, 
by using the same notation as in Figs~\ref{fig:F_tauM} and~\ref{fig:F_tauMcompact}, 
signal $y(t)$ can be written as 
\begin{equation}
y(t) = \prod_{i=1}^d f^i(t) = \prod_{i=1}^d (f_1^i(t)+f_2^i(t)+f_3^i(t)+f_4^i(t))
=\sum_{m\in \{1,2,3,4\}^d} f_{m_1}^1(t)\cdot \ldots \cdot f_{m_d}^d(t).
\end{equation}
Now it is clear that product in the last sum in (3) 
can be written as
\begin{equation}
f_{m_1}^1(t)\cdot \ldots \cdot f_{m_d}^d(t)= \prod_{i=1}^d e_{m_i}^i(t) \cdot 
\prod_{k\in S_1(m)\cup S_2(m)}\cos(t-\tau_k) \cdot
\prod_{l\in S_3(m)\cup S_4(m)}\sin(t-\tau_l),
\end{equation}
where $e_{m_i}^i(t)$ equals $f_{m_i}^i(t)/ \cos(t-\tau_i)$ for $m_i=1,2$, 
or $f_{m_i}^i(t)/ \sin(t-\tau_i)$ otherwise. Since our goal is to find a 
transfer function from $w[n]$ to $v[n]$, it is more convenient to 
express the above products of cosines and sines as sums of 
complex exponentials, i.e.
\[
\prod_{k\in S_1(m)\cup S_2(m)}\cos(t-\tau_k)=\\
 \frac{1}{2^{N_1(m)}}\sum_{r\in\{-1,1\}^{N_1(m)}}e^{j\omega_c\bar \sigma(r)t}
\cdot e^{j\omega_c(r,\mathcal{P}_m^1\tau)},
\]
\[
\prod_{l\in S_3(m)\cup S_4(m)}\sin(t-\tau_l)= \\
\frac{1}{(2j)^{N_2(m)}}\sum_{r\in\{-1,1\}^{N_2(m)}}(-1)^{\prod_{i=1}^{N_2(m)}r_i}
\cdot e^{j\omega_c\bar\sigma(r)t}\cdot e^{j\omega_c(r,\mathcal{P}_m^2\tau)}.
\]
Signals $e_{m_i}^i(t)$ are obtained by applying pulse amplitude 
modulation with $p_1(t)$ or $p_2(t)$ on in-phase or quadrature 
components of the input signal (or their delayed counterparts). 
Now their product can be written as
\begin{equation}
\prod_{i=1}^d e_{m_i}^i(t) = \sum_{n=-\infty}^\infty 
i[n]^{|S_1(m)|}i[n-1]^{|S_2(m)|}\cdot \\
\cdot q[n]^{|S_3(m)|}q[n-1]^{|S_4(m)|}p_m(t-nT).
\end{equation}
If we denote this product by $e_m(t)$, we can write (4) as
\begin{equation}
f_{m_1}^1(t)\cdot \ldots \cdot f_{m_d}^d(t) = 
e_m(t)\cdot\frac{ (-j)^{N_2(m)}}{2^d}\cdot
\sum_{r\in\{-1,1\}^d}(-1)^{\pi_m(r)}\cdot 
e^{j\omega\bar \sigma(r)t} \cdot e^{-j\omega_c(r,\tau)}.
\end{equation}
Finally from (3),(5) and (6) it follows that the output $v[n]$ 
is equal to
\[v[n]=\sum_{m\in \{1,2,3,4\}^d}f_{m}[n]*h_{m,\tau}[n],\]
where 
\[f_m[n]=i[n-1]^{|S_1(m)|} \cdot i[n]^{|S_2(m)|} \cdot 
q[n-1]^{|S_3(m)|} \cdot q[n]^{|S_4(m)|},\]
and Fourier transforms of impulse responses $h_{m,\tau}[n]$ 
are given by
\[
H_{m,\tau}(e^{j\Omega})=\sum_{m\in \{1,2,3,4\}^d}\frac{ (-j)^{N_2(m)}}{2^d} 
\sum_{r\in\{-1,1\}^d}(-1)^{\pi_m(r)}\cdot 
P_m\left(j\frac{\Omega}{T}-j\omega_c \sigma(r)\right) \cdot e^{-j\omega_c(r,\tau)}.
\]
This concludes the proof of Lemma 2.2.
\end{proof}
\vskip2mm
In Lemma 2.2 we assumed that $\tau_i\in[0,T),\ \forall i\in[1:d]$, 
but in general $\tau_i$ can take any positive real value depending 
on the depth of (2), i.e. vector $k$ associated to $\tau$ is not 
necessarily zero vector. Now assume that $\tau = kT+\bar {\tau}$, where 
$\bar{\tau}\in[0,T)^d$. The input/output relation for system 
$\mathbf{DHF_{\tau}M}$ readily follows from Lemma 2.2, and we have
\[ v[n]=\sum_{m\in\{1,2,3,4\}^d}f_m[n]*h_m[n],\]
where signals $f_m[n]$ are given by
\[
f_m[n]=\prod_{i\in S_1(m)}i[n-k_i-1] \cdot \prod_{i\in S_2(m)} i[n-k_i] \cdot
\prod_{i\in S_3(m)} q[n-k_i-1] \cdot \prod_{i\in S_4(m)}q[n-k_i],
\]
and unit sample responses $h_{m,\tau}[n]$ have the following Fourier transforms
\[
H_{m,\tau}(e^{j\Omega})=\frac{ (-j)^{N_2(m)}}{2^d} 
\sum_{r\in\{-1,1\}^d}(-1)^{\pi_m(r)}\cdot
P_m\left(j\frac{\Omega}{T}-j\omega_c \sigma(r)\right) \cdot e^{-j\omega_c(r,\tau)}.
\]
\end{proof}


\section{Simulation Results}
\begin{figure}[H]
\setlength{\unitlength}{0.12cm}
\begin{center}\begin{picture}(64,36)(8,36)
\put(24,64){\framebox(11,8){$\mathbf V_0$}}
\put(24,50){\framebox(11,8){$\mathbf V_1$}}
\put(24,36){\framebox(11,8){$\mathbf V_2$}}
\put(35,68){\vector(1,0){12}}
\put(47,64){\framebox(11,8){$\mathbf H_0$}}
\put(35,54){\vector(1,0){12}}
\put(47,50){\framebox(11,8){$\mathbf H_1$}}
\put(35,40){\vector(1,0){12}}
\put(47,36){\framebox(11,8){$\mathbf H_2$}}
\put(20,68){\vector(1,0){4}}
\put(20,68){\line(0,-1){28}}
\put(20,40){\vector(1,0){4}}
\put(20,54){\vector(1,0){4}}
\put(8,54){\vector(1,0){12}}
\put(58,54){\vector(1,0){4}}
\put(64,54){\circle{4}}
\put(58,68){\line(1,0){4}}
\put(62,68){\line(1,0){2}}
\put(64,68){\vector(0,-1){12}}
\put(58,40){\line(1,0){6}}
\put(64,40){\vector(0,1){12}}
\put(66,54){\vector(1,0){6}}
\put(64,55){\line(0,-1){2}}
\put(63,54){\line(1,0){2}}
\put(10,56){$w[n]$}
\put(38,56){$g_1[n]$}
\put(38,70){$g_0[n]$}
\put(38,42){$g_2[n]$}
\put(69,56){$v[n]$}
\end{picture}\end{center}
\caption{Proposed compensator structure}
\label{fig:Compensator_structure}
\end{figure}
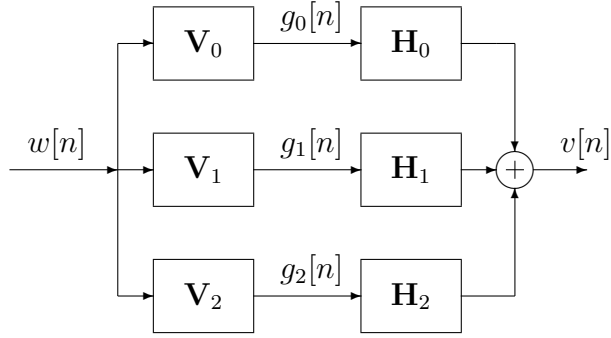

In this section, through MATLAB simulations, we illustrate performance of the 
proposed compensator structure. We compare this structure with some standard 
compensator structures, together with ideal compensator, and show 
that it closely resembles dynamics of ideal compensator, thus achieving very good 
compensation performance. \\
The underlying system {\bf S} is given in Figure~\ref{fig:whole_chain}, with the 
distortion subsystem {\bf F} given by
\begin{equation} \label{eq:distortion}
(Fx)(t)=x(t)-\delta\cdot x(t-\tau_1)x(t-\tau_2)x(t-\tau_3),
\end{equation}
where $0\le \tau_1 \le \tau_2 \le \tau_3 \le T$, with $T$ sampling time, and 
$\delta>0$ parameter specifying magnitude of distortion $\bf \Delta$ in 
${\bf S}={\bf I}+{\bf \Delta}$. We assume that parameter $\delta$ is relatively 
small, in particular $\delta \in (0,0.2)$, so that the inverse $\mathbf S^{-1}$ of {\bf S} can be well 
approximated by $2\mathbf I - \mathbf S$. 
Then our goal is to build compensator $\mathbf C=\mathbf S^{-1}$ with 
different structures, and compare their performance, which is measured as 
output Error Vector Magnitude (EVM)  \cite{VuoRah2003}. EVM, for an input 
$u$ and output $\hat u$, is defined as
\[\textrm{EVM(dB)} =20\log_{10} \left(\frac{||u-\hat u||_2}{||u||_2}\right).\]
Analytical results from the previous section suggest that the compensator structure 
should be of the form depicted in Figure~\ref{fig:compensator}.
It is easy to see from the proof of Theorem 1.1, that transfer functions in {\bf L}, 
from each nonlinear component $g_k[n]$ of $g[n]$, to the output $v[n]$, are 
smooth functions, hence can be well approximated by low order 
polynomials in $\Omega$. In this example we choose second order polynomial 
approximation of components of {\bf L}. This observation, together with the true 
structure of {\bf S}, suggests that compensator {\bf C} should be fit
within a family of models with structure shown on the block diagram in 
Fig~\ref{fig:Compensator_structure}.

\noindent Subsystems $\mathbf H_i, i=1,2,3$, are LTI systems, with 
transfer functions $H_i$ given by
\[H_0(e^{j\Omega})=1, H_1(e^{j\Omega})=j\Omega, 
H_2(e^{j\Omega})=\Omega^2, \forall \Omega \in [-\pi,\pi].\]
Nonlinear subsystems $\mathbf V_i$ are modeled as third order Volterra 
series, with memory $m=1$, i.e.

\[  ({\mathbf V_j}w)[n]=\sum_{(\alpha(k),\beta(k))}c^j_k
\prod_{l=0}^{1}i[n-l]^{\alpha_l(k)}~
\prod_{l=0}^{1}q[n-l]^{\beta_l(k)},\]
\[\alpha_l(k),\beta_l(k)\in\mathbb Z_+,\qquad
 \sum_{l=0}^1\alpha_l(k)+\sum_{l=0}^1\beta_l(k)\le 3,\]
where $i[n]=\text Re~w[n]$ and $q[n]=\text Im~w[n]$, and 
$(\alpha(k),\beta(k))=(\alpha_0(k),\alpha_1(k),\beta_0(k),\beta_1(k))$.

We compare performance of this compensator with the widely used one 
obtained by utilizing simple Volterra series structure \cite{VuoRah2003}:
\[  ({\mathbf C}w)[n]=\sum_{(\alpha(k),\beta(k))}c_k
\prod_{l=-m_1}^{m_2}i[n-l]^{\alpha_l(k)}~
\prod_{l=-m_1}^{m_2}q[n-l]^{\beta_l(k)},\]
\[\alpha_l(k),\beta_l(k)\in\mathbb Z_+,\qquad
 \sum_{l=-m_1}^{m_2}\alpha_l(k)+\sum_{l=-m_1}^{m_2}\beta_l(k)\le d.\]
Parameters which could be 
varied in this case are forward and backward memory depth $m_1$ and $m_2$, 
respectively, and degree $d$ of this model. We consider three cases 
for different sets of parameter values: 
\begin{itemize}
\item Case 1: $m_1=0,~m_2=2,~d=5$
\item Case 2: $m_1=0,~m_2=4,~d=5$
\item Case 3: $m_1=2,~m_2=2,~d=5$
\end{itemize}

\begin{table}[h]
\caption{Number of coefficients $c_k$ being optimized for different compensator models}
\label{table}
\begin{center}
\begin{tabular}{|c||c|c|}
\hline
Model & \# of $c_k$ & \# of significant $c_k$\\
\hline\hline
New structure & 210 & 141\\
\hline
Volterra 1 & 924 & 177\\
\hline
Volterra 2 & 6006 & 2058\\
\hline
Volterra 3 & 6006 & 1935\\
\hline
\end{tabular}
\end{center}
\end{table}

After fixing compensator structure, coefficients $c_k$ are obtained by 
applying straightforward least squares optimization. \\
We should emphasize here that fitting has to be done 
for both real and imaginary part of $v[n]$, thus the actual compensator structure 
is twice that depicted in Figure~\ref{fig:Compensator_structure}. \\

\begin{figure}
\center
\includegraphics[scale=0.9]{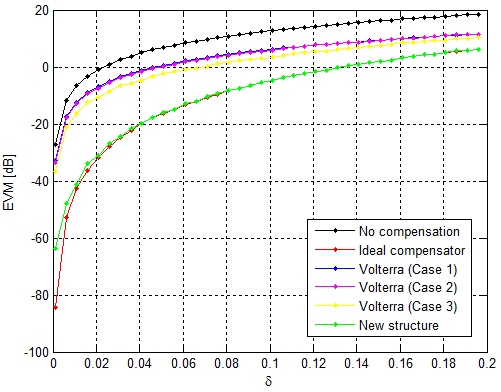}
\caption{Output EVM for different compensator structures}
\label{fig:EVM_comparison}
\end{figure}

Simulation parameters for system {\bf S} are as follows: symbol rate 
$f_{symb}=2\text{MHz}$, carrier frequency $f_c=20\text{MHz}$, 
with 64QAM input symbol sequence. Nonlinear distortion subsystem {\bf F} 
of {\bf S}, used in simulation, is defined in \eqref{eq:distortion}, where the delays 
$\tau_1,\tau_2,\tau_3$ are given by the vector $\tau = [0.2T\quad 0.3T \quad 0.4T]$, 
with $T=1/f_{symb}$. Digital simulation of the continuous part of {\bf S} was done 
by representing continuous signals by their discrete counterparts, obtained by 
sampling with high sampling rate $f_s=1000\cdot f_{symb}$. 
As input to {\bf S}, we assume periodic 64QAM symbol sequence, with period 
$N_{symb}=4096$. This period length is used for generating input/output data 
for fitting coefficients $c_k$, as well as generating input/output data for performance 
validation.  
\\
In Figure~\ref{fig:EVM_comparison} we present EVM obtained for different compensator 
structures, as well as output EVM with no compensation, and case with ideal compensator 
$\mathbf C=\mathbf S^{-1}\approx 2\mathbf I-\mathbf S$. As can be seen from 
Figure~\ref{fig:EVM_comparison}, compensator fitted using the proposed structure 
in Figure~\ref{fig:Compensator_structure}
outperforms other compensators, and gives output EVM almost identical to the ideal
compensator. This result was to be expected, since model in 
Figure~\ref{fig:Compensator_structure} approximates the original system {\bf S} 
very closely, and thus is capable of approximating system $2\mathbf I-\mathbf S$ 
closely as well. This is not the case for compensators modeled with simple Volterra 
series, due to inherently long (or more precisely infinite) memory introduced 
by the LTI part of {\bf S}. Even if we use noncausal Volterra series model 
(i.e. $m_1\not=0$), which 
is expected to capture true dynamics better, we are still unable to get good fitting 
of the system {\bf S}, and consequently of the compensator 
$\mathbf C\approx 2\mathbf I-\mathbf S$.\\
Advantage of the proposed compensator structure is not only in better 
compensation performance, but also in that it achieves better performance 
with much more efficient strucuture. That is, we need far less 
coefficients in order to represent nonlinear part of the compensator, in 
both least squares optimization and actual implementation (Table~\ref{table}).  
In Table~\ref{table} we can see a comparison in the number of coefficients 
between different compensator structures, for nonlinear subsystem parameter 
value $\delta = 0.02$.  Data in the first column is number of coefficients 
(i.e. basis elements) needed for general Volterra model, i.e. coefficients 
which are optimized by least squares. The second column shows actual number 
of coefficients 
used to build compensator. Least squares optimization yields many nonzero 
coefficients, but only subset of those are considered 
significant and thus used in actual compensator implementation. 
Coefficient is considered significant if its value falls above 
a certain treshold $t$, where $t$ is chosen such that increas in EVM after zeroing 
nonsignificant coefficients is not larger than 1\% of the best achievable EVM 
(i.e. when all basis elements are used for building compensator). From 
Table~\ref{table} we can see that for case 3 Volterra structure, 10 times more 
coefficients are needed in order to implement compensator, 
than in the case of our proposed structure. And even when such a large number of 
coefficients is used, its performance is still below the one achieved by this new 
compensator model.


\section{Discussion}
The potential significance of the result presented in this paper lies in
revealing a special structure of a digital pre-distortion compensator which appears to
be both
necessary and sufficient to match the discrete time dynamics resulting from combining
modulation and demodulation with a dynamic non-linearity in continuous time.
The "necessity" somewhat relies on the input signal $u$ having "full" spectrum.
While, theoretically, the baseband signal $u$ is supposed to be shaped so that only a
lower DT frequency spectrum of it remains significant, a practical implementation of
amplitude-phase modulation will frequently employ the a  
signal component separation approach, such as LINC \cite{Cox1974}, where the
low-pass signal $u$ is decomposed into two components of constant amplitude,
$u=u_1+u_2$, $|u_1[n]|\equiv|u_2[n]|=\text{const}$, after which the components
$u_i$ are fed into two separate modulators, to produce continuous time
outputs $y_1,y_2$, to be combined into a single output $y=y_1+y_2$. Even when
$u$ is band-limited, the resulting components $u_1,u_2$ are not, and the full range of
modulator's nonlinearity is likely to be engaged when producing $y_1$ and $y_2$.




\section*{Acknowledgment}
The authors are grateful to Dr. Yehuda Avniel for bringing researchers from
vastly different backgrounds to work together on the tasks that led to the writing of
this paper.


\end{document}